\documentclass[usenatbib]{mn2e}
\usepackage{graphicx}
\usepackage{amsmath}
\usepackage{nicefrac}
\usepackage{paralist}
\usepackage{subfig}
\usepackage{url}

\title[Real-time, fast radio transient searches with GPU
  de-dispersion]{Real-time, fast radio transient searches with GPU
  de-dispersion}

\author[A. Magro et al.]  {A. ~Magro$^{1}$, A. ~Karastergiou$^{2}$,
  S. ~Salvini$^{3}$, B. ~Mort$^{3}$, F. ~Dulwich$^{3}$ and \newauthor
  K. ~Zarb ~Adami $^{1,2}$
\\
$^{1}$Department of Physics, University of Malta, Msida MSD 2080,
  Malta \\ 
$^{2}$Astrophyics, University of Oxford, Denys Wilkinson Building,
  Keble Road, Oxford OX1 3RH, UK \\ 
$^{3}$Oxford e-Research Center, University of Oxford, 7, Keble Road,
  Oxford OX1 3QG, UK}

\date{Released 2011 Xxxxx XX}
\pagerange{\pageref{firstpage}--\pageref{lastpage}}

\begin{document}

\maketitle
\label{firstpage}

\begin{abstract}

The identification, and subsequent discovery, of fast radio transients
through blind-search surveys requires a large amount of processing
power, in worst cases scaling as $\mathcal{O}(N^3)$. For this reason,
survey data are generally processed offline, using high-performance
computing architectures or hardware-based designs. In recent years,
graphics processing units have been extensively used for numerical
analysis and scientific simulations, especially after the introduction
of new high-level application programming interfaces. Here we show how
GPUs can be used for fast transient discovery in real-time. We present
a solution to the problem of de-dispersion, providing performance
comparisons with a typical computing machine and traditional pulsar
processing software. We describe the architecture of a real-time,
GPU-based transient search machine. In terms of performance, our GPU
solution provides a speed-up factor of between 50 and 200, depending
on the parameters of the search.

\end{abstract}

\section{Introduction}
\label{introductionSection}

One of the most ambitious experiments of next-generation
radio-telescopes, such as the newly inaugurated Low Frequency Array
(LOFAR) and the future Square Kilometre Array (SKA) is to explore the
nature of the dynamic radio sky at timescales ranging from nanoseconds
to years. Surveys for very fast radio transients necessarily generate
huge amounts of data, making storage and off-line processing an
unattractive solution. On the other hand, real-time processing offers
the possibility to react as fast as possible and conduct follow-up
observations across the electromagnetic spectrum and even, for some
events, with gravitational wave detectors. Here we will consider the
case where time-series data (tied-array, or beam-formed data for array
telescopes mentioned above) are processed to extract astrophysical
radio bursts of short duration. The multitude of potential
astrophysical events that may produce such transients, ranging from
individual pulses from neutron stars to Lorimer bursts
\citep{lorimer2007}, are discussed in \cite{cm04}.

In this paper we will mostly be concerned with de-dispersion. This
refers to a family of techniques employed to reverse the
frequency-dependent refractive effect of the interstellar medium (ISM)
on the radio signals passing through it. We also use interstellar
scattering to constrain the parameter space of a given search - see
\citep[chapter 4]{LorimerKramer2005} for more details on these
effects. From pulsar studies, it is well determined that propagation
through the ISM obeys the cold plasma dispersion law, where the
time-delay between two frequencies $f_1$ and $f_2$ is given by the
quadratic relation
\begin{equation}
\label{dispRelationshipEquation}
 \Delta t \simeq 4.15 \times 10^6 \mbox{ ms} \times \left( f^{-2}_1 - f^{-2}_2 \right) \times \mbox{DM}
\end{equation}
where DM is the dispersion measure in $\text{pc cm}^{-3}$ and $f_1$
and $f_2$ are in MHz. Astrophysical objects have a particular DM value
associated with them, which depends on the total amount of free
electrons along the line of sight and, therefore, on the distance to
the object from Earth.

Removing the effect of dispersion from astrophysical data is typically
done in two ways, depending on the type of data and the requirements
of the experiment. For total power filterbank data, where the spectral
bandwidth of the observation is typically split into a number of
narrow frequency channels, de-dispersion consists of a relative shift
in time of all frequency channels according to equation
\ref{dispRelationshipEquation}. This is known as incoherent
de-dispersion as it is performed on incoherent data. For baseband
data, de-dispersion can be done by convolution of the observed voltage
data with the inverse of the transfer function of the ISM. This is
known as coherent de-dispersion; it is more accurate in terms of
recovering the intrinsic shape of the astrophysical signal but much
more demanding in computational power. In this paper, we deal with
incoherent de-dispersion, applied to total power data.

De-dispersion is an expensive operation, scaling as
$\mathcal{O}\left(N^2\right)$ for brute-force algorithms, with more
optimized techniques diminishing this to
$\mathcal{O}\left(Nlog(N)\right)$. For any blind, fast transient
search, de-dispersion needs to be performed over a range of DM values,
practically multiplying the number of operations required by the
number of DM values in the search. Typically, thousands of finely
spaced DM values need to be considered, so the entire operation
becomes extremely compute-intensive.  In the past, searches for fast
transients, such as single-pulse pulsar searches, have been performed
offline on archival data, using standard processing tools on large
computing clusters or supercomputers. The power of such searches is
highlighted by the discovery of Rotating Radio Transients (RRATs)
\citep{mll+06}.

Here we discuss how such searches can be performed online by using
standard off-the-shelf general purpose graphics processing units
(GPUs).  With the introduction of new high level application
programming interfaces (APIs) in recent years, such as CUDA
\citep{cudaOnline}, these devices can easily be used for offloading
data-parallel, processing-intensive algorithms from the CPU.  The idea
of performing transient-related signal processing on GPUs is not a new
one. For example, Allal et al. (2009) have created a fully working,
real-time, GPU-based coherent de-dispersion setup at the Nan\c{c}ay
Radio Observatory \nocite{Allal2009}. The digital signal processing
library for pulsar astronomy written by van Straten and Bailes
(2010)\nocite{VanStraten2010} can also use GPUs for performing
coherent de-dispersion, folding and detection.

\nocite{Barsdell2010}Barsdell, Barnes \& Fluke (2010) analyse various
foundation algorithms which are used in astronomy and try and
determine whether they can be implemented in a massively-parallel
computing environment. One of the algorithms they analyse is the
brute-force de-dispersion algorithm, and they state that this would
likely perform to high efficiency in such an environment, whilst
stating that optimal arithmetic intensity is unlikely to be achieved
without a detailed analysis of the algorithm's memory access
patterns. The incoherent de-dispersion algorithm has indeed been
implemented using CUDA as a test-case for the ASKAP CRAFT project (see
Macquart et al. and Dodson et
al. 2010)\nocite{Macquart2010,Dodson2010}.

\section{Blind fast transient search parameters}
\label{parametersSection}

Fast transient surveys rely on a number of search parameters, which
depend on the characteristics of the instruments being used and the
characteristics of the target astrophysical signals \citep{cor08}.
For example, a survey for dispersed fast transients can be designed
using the following input parameters:
\begin{inparaenum}
 \item the \textit{center frequency} at which the observation is conducted
 \item the frequency \textit{bandwidth}
 \item the \textit{number of frequency channels and channel bandwidth}
 \item the \textit{sampling rate} at which digital data are available
 \item the expected \textit{signal width}, which depends on the
   science case
 \item the acceptable \textit{signal-to-noise (S/N) level} of a
   detected signal.
\end{inparaenum}

Dispersion and scattering are dependent on the frequency and bandwidth
at which the survey is being conducted, and help define the boundaries
of a survey as follows:
\begin{description}
 \item[{\bf The maximum DM value}] ($DM_{\rm max}$) can be chosen
   according to the maximum DM value at which there is a justified
   expectation to discover sources. This may be related to the
   Galactic coordinates of the survey, the capabilities of the
   hardware and also considerations related to interstellar
   scattering. Scattering has the effect of reducing the peak S/N of a
   signal, and is related to DM via an empirical relationship
   \citep[see][]{Bhat2004}.

 \item[{\bf The number of frequency channels}] depends on the maximum
   acceptable channel bandwidth. This can be chosen by considering the
   dispersion smearing within each channel for the larger DMs in the
   searched DM range. Depending on the format of the data handed down
   by the telescope, generating narrower frequency channels may be
   desirable and necessary.

 \item[{\bf The de-dispersion step}], which is the unit of
   discretisation of the DM range, is chosen according to the width of
   the target astrophysical signals and the permissible S/N loss that
   occurs when de-dispersing at a slightly erroneous DM.
\end{description}

De-dispersion itself can be performed by a brute force algorithm,
introducing time shifts to every channel, or using faster,
approximative de-dispersion algorithms. Subband de-dispersion is a
technique used by PRESTO\footnote{PRESTO is a pulsar search and analysis 
software developed by Scott Ransom} \citep{Ransom2001}, and it
relies on the fact that adjacent DM values often reuse the same time
samples to create the de-dispersed time series. The entire band is
split into a number of groups of channels, or subbands. Each subband
is de-dispersed according to a set of coarsely spaced DM values and
collapsed into a single frequency channel, representative of the
band. The reduced number of pre-processed channels are then
de-dispersed using a much finer DM step, resulting in de-dispersed
time-series for all DMs within the search range.  This technique
results in a slight sensitivity loss, but greatly decreases the
processing time. Subband de-dispersion requires additional parameters,
which in PRESTO are provided as a survey plan. This dictates how the
DM range is split into \textit{passes}, where each pass will bin the
data using a different binning factor (the larger the DM, the more the
signal will be smeared in time, thus we can reduce the computational
cost by averaging samples). For each pass the DM range is partitioned,
each range being $\Delta Sub_{DM}$ apart.  This value dictates the
subband DM step between consecutive nominal DM values. The DM step is
then used to split each of these partitions.

\begin{table}
  \begin{tabular}{c | c*{6}{c}}
  Pass     & Low DM & High DM & $\Delta$DM & Bin & $\Delta$ Sub$_{\text{DM}}$ \\
           & ($\text{pc cm}^{-3}$) & ($\text{pc cm}^{-3}$) & ($\text{pc cm}^{-3}$) & & ($\text{pc cm}^{-3}$) \\
  \hline
  1        & 0.00   & 53.46   & 0.03         & 1   & 0.66              \\
  2        & 53.46  & 88.26   & 0.05         & 2   & 1.2               \\
  3        & 88.26  & 150.66  & 0.10         & 4   & 2.4               \\
  \end{tabular}
   \centering
   \caption{An example of a subband de-dispersion survey plan. $\Delta
     Sub_{DM}$ refers to the DM step between two successive nominal DM
     values, while $\Delta DM$ is the finer DM step used for creating
     the de-dispersed time-series around a particular nominal DM
     value. Bin refers to the binning coefficient for a particular
     pass.}
   \label{surveyPlanTableExample}
\end{table}

Table \ref{surveyPlanTableExample} provides an example of such a
survey plan for a DM range of 0 - 150.66 as produced by DDplan, a
script included with PRESTO, which generates the survey parameters by
trying to minimize the smearing induced by the splitting of the
bandwidth into subbands. This smearing is the additive effect of the
smearing over each channel, each subband, the full bandwidth and the
sampling rate (assuming the worst-case DM error).

\section{GPU-based de-dispersion}
\label{dedisprsionSection}

\begin{figure}
\centering
\includegraphics[width=80mm]{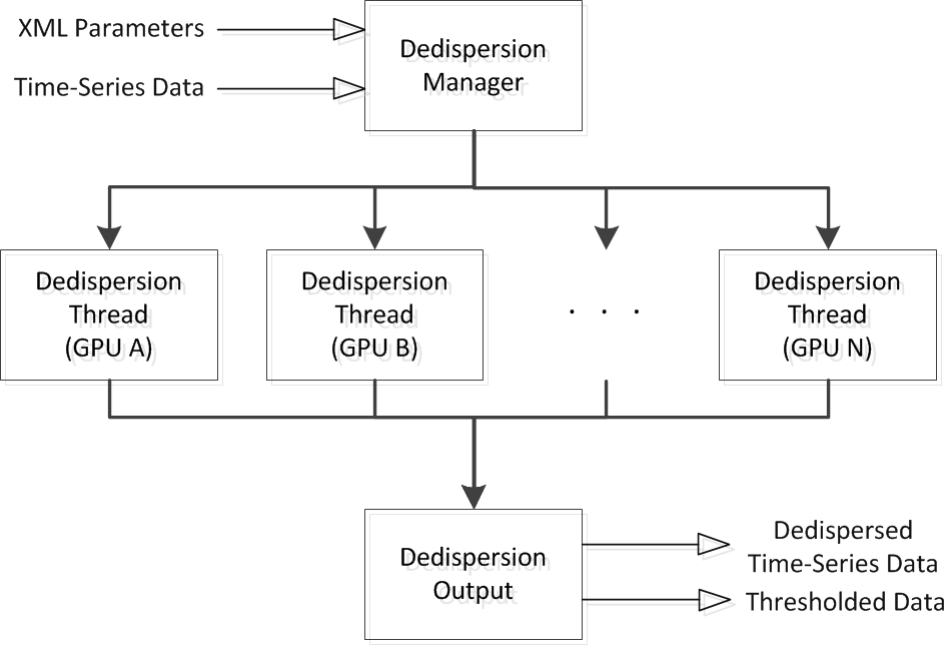}
\caption[High-level thread hierarchy in MDSM] 
        {The high-level thread hierarchy. The de-dispersion manager initializes the entire system and then takes care
         of reading input data. A de-dispersion thread per GPU is created which handles all memory copying and kernel launches on
         that GPU. The de-dispersion output thread then receives the de-dispersed time series from all the GPUs and performs a simple
         burst search.}
\label{mdsmThreadHierarchyFigure}
\end{figure}

Using the CUDA API, we have implemented incoherent de-dispersion,
which works on any CUDA-capable GPUs attached to a host computer. Our
modular code parallelizes the host and GPU execution by using multiple
threads during the input, processing and output parts. While the input
thread is reading data, the GPU is busy de-dispersing a previously
read buffer and the output thread is post-processing a de-dispersed
time-series. This threaded setup is depicted in figure
\ref{mdsmThreadHierarchyFigure}.

The de-dispersion manager is the main thread which takes care of
initialising and synchronizing the rest of the application.  The input
thread handles all CUDA-related calls, and an instance for each
attached GPU is created.  The DM range is split among these threads,
such that each will process the same input buffer for different DM
values.  The de-dispersed output from GPU memory is then copied to
host memory, to an address which is shifted appropriately to
accommodate all the input threads.

The output thread constructs the de-dispersed time-series and outputs
the results to a data file. It may also be responsible for event
detection. The simplest approach is then to calculate the mean ($\mu$)
and standard deviation ($\sigma$) for the entire processed buffer, and
use these values to apply a threshold at a particular multiple of the
standard deviation ($n\sigma$). All values above the threshold are
output to file as a list of triplets of the form
$\left(time,\;DM,\;intensity\right)$.

\begin{figure*}
\centering
\includegraphics[width=120mm]{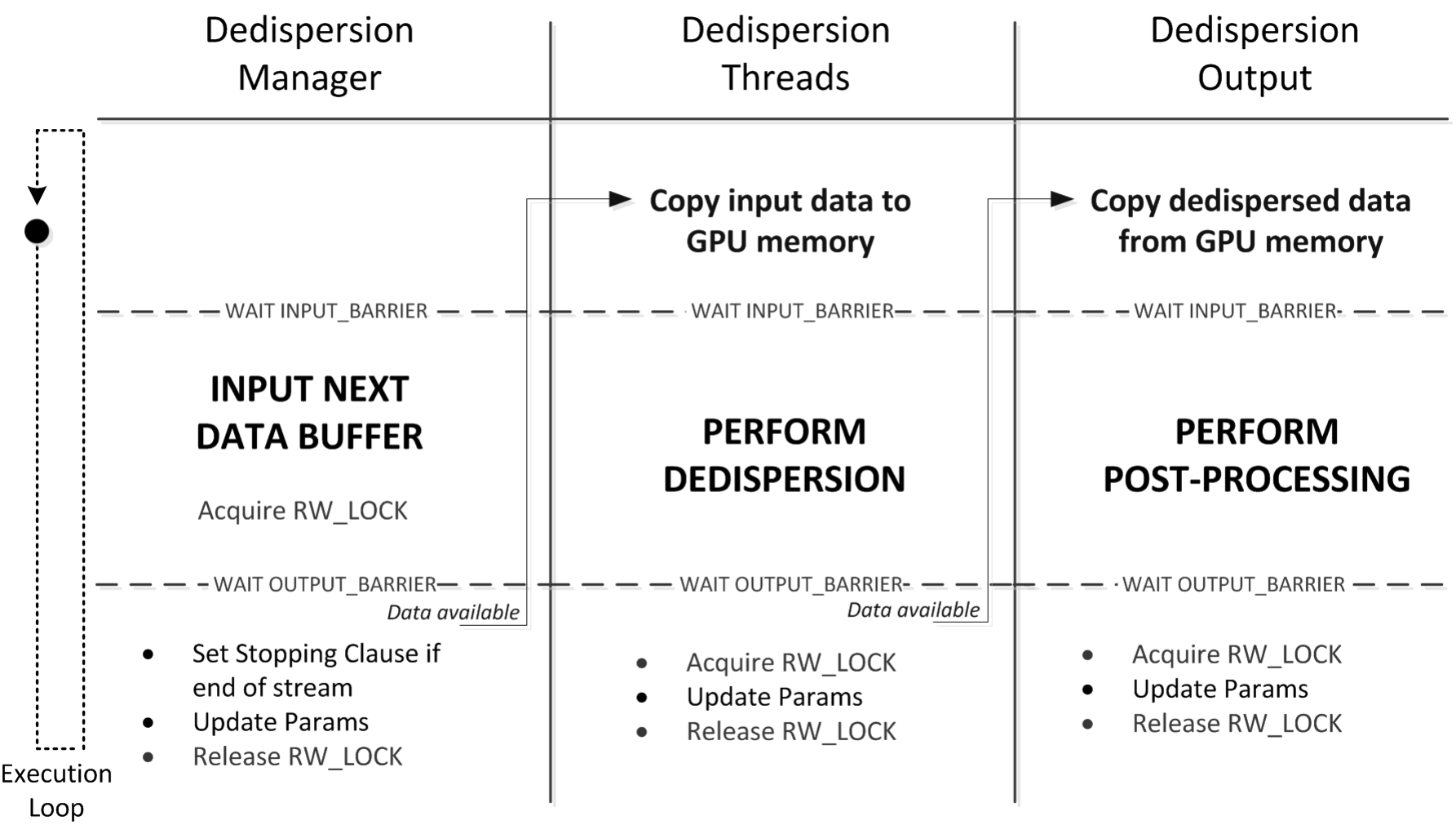}
\caption[Thread, synchronization and processing flow in MDSM]
{Each thread has three main stages: the input, processing and output
  stages. Data has to flow from one thread type to the next, so
  synchronization objects have to be used to make sure that no data is
  overwritten or re-processed.  Barrier and RW locks are used to
  control access to critical sections. See main text for a more
  detailed explanation.}
\label{mdsmFlowFigure}
\end{figure*}

Currently a homogeneous system is assumed, and no load-balancing
between the devices is performed. Each thread is split into three
conceptual ``processing stages'', which are guarded by several
thread-synchronization mechanisms. This setup is shown in figure
\ref{mdsmFlowFigure}. The three stages are:
\begin{inparaenum}[\itshape (i)]
\item the input section, where the thread inputs data to be processed
\item the processing section, which contains the de-dispersion kernel,
  is the main section in the thread and the part which takes the
  longest to complete \item the output section, where the processed
  buffer is output and
  made available to the next thread and any parameter updates are
  performed.
\end{inparaenum}

The two aforementioned de-dispersion techniques have been implemented,
namely brute force and subband de-dispersion, which between them have
some common elements:
\begin{description}
\item[{\bf Maximum shift:}] In a buffer containing $n_{\rm samp}$
  samples to process with a non-zero DM value, each channel will
  require a certain shift, with the lower frequency channels requiring
  the greatest shift. Assuming this shift is of $s$ samples, and
  $n_{\rm samp}$ samples need to be de-dispersed, we require $n_{\rm
    samp}+s$ samples to be available, where $s$ is dependent on the DM
  value. Since the GPU will be de-dispersing for $n_{\rm dms}$ DM
  values at any one time, the amount of extra samples need to cater
  for the maximum DM value.  For reference, we will use the term
  \textit{maxshift} ($m_{\rm shift}$) for this shift, which can be
  calculated by manipulating equation \ref{dispRelationshipEquation}:
  \begin{equation}
    m_{\rm shift} = \frac{8.3\times 10^6ms\times \Delta f\times f^{-3}\times DM_{\rm max}}{t_{\rm samp}}
  \end{equation}
  where $t_{\rm samp}$ is in ms, $DM_{\rm max}$ is the maximum DM
  value processed on the GPU and $f$ and $\Delta f$ are in MHz.
  
\item[{\bf Processable Samples:}] Data transfers between the GPU and
  CPU are inefficient. For this reason the input buffer should fill up
  as much of the GPU's memory as possible, leaving enough space to
  store the maxshift and output buffer. The simplest way to calculate
  the number of samples which fit in memory is:
  \begin{equation} 
    n_{\rm samp} = \frac{memory-\left(m_{\rm shift} \times n_{\rm chans}\right)}{n_{\rm dms}+n_{\rm chans}}
  \end{equation}
  The aim is to keep all the data within the GPU memory and perform
  all processing there, so if there are a number of operations to be
  performed (such as binning), $n_{\rm samp}$ must accommodate these
  as well.
  
\item[{\bf Shifts:}] Each channel requires a different amount of
  shift, for each DM. For an input buffer with $n_{\rm chans}$
  channels, when de-dispersing for $n_{\rm dms}$ DM values, the
  required data structure has a size of $n_{\rm dms}\times n_{\rm
    chans}$ values. For the amount of channels and DM values required
  for most situations, such a data structure will not fit in constant
  memory, and would greatly reduce the execution speed if they were
  calculated for each block, for each DM. Storing these values in
  global memory would slow the kernel down.  To counter this, the
  calculation is split in two parts, with the first part performed on
  the CPU:
  \begin{equation}
    t_{chan} = 4.15 \times 10^3 \times \left(f_1^{-2}-f_2^{-2}\right)
  \end{equation}
  where $f_1$ and $f_2$ are in MHz. This gives us a DM-independent
  shift for each frequency channel (i.e. the dispersion delay per unit
  DM), resulting in a data structure of size $n_{\rm chans}$, which in
  normal circumstances will fit in constant memory. The second part of
  the calculation is then performed on the GPU:
  \begin{equation}
    t_{DM} = \frac{t_{chan} \times DM}{t_{\rm samp}}
  \end{equation}
  The division by $t_{\rm samp}$ could also be performed on the CPU,
  but this would result in rounding errors when casting the result to 
  single-point precision (the value on which the GPU would operate) 
  since $t_{chan}$ is usually very small; it is more efficient to perform 
  the calculation on the GPU rather than using double precision throughout.
\end{description}

\subsection{Brute-force incoherent de-dispersion}
\label{bruteForceGpu}

The brute-force algorithm is the simplest and most accurate to
implement for de-dispersion of incoherent data, but is also the least
efficient processing wise. Assuming $N_s$ samples, each with $N_c$
channels, and de-dispersing for $N_{DM}$ DM values, the algorithmic
time complexity of the brute force algorithm is
$\mathcal{O}\left(N_s\times N_c \times
  N_{DM}\right)$. $N_s$ can be seen as an infinite stream of samples,
while $N_c$ and $N_{DM}$ will usually have a similar value, resulting
in approximately $N^2$ operations for every input sample.

According to \citep{Barsdell2010} there are three main ways in which this algorithm can be parallelized:
\begin{inparaenum}[\itshape (a)]
\item $N_s$ parallel threads each compute the sum of a single input time sample for every channel sequentially
\item $N_c$ parallel threads cooperate to sum each sample in turn
\item $N_s\times N_c$ parallel threads cooperate to complete the entire computation in parallel.
\end{inparaenum}
The current implementation uses a variant of scheme (a), where each
thread sums up the input for a single time sample. Due to the large
number of samples which can fit in GPU memory, each thread will end up
processing more than one sample. A way to envisage this is to imagine the
CUDA grid as a sliding window which moves along the input samples at
discrete intervals equal to the total number of threads in one row. At
each grid position, threads are assigned to their respective samples.

The kernel can process any number of DM values concurrently, and this
is done by creating a two-dimensional grid, where each row is assigned
a different DM value for de-dispersion. The output of $N_{DM}$
time-series, each with $N_s$ samples, is output to the output thread
for post-processing.

This kernel is not very compute-intensive, performing less than ten
floating point operations per global memory read. This makes the
de-dispersion algorithm memory-limited. For this reason, depending on
the way the data are read from the input device, a corner-turn (matrix
transpose) might be required in order to store the data in channel
order. With this memory setup, and having each thread process one
sample for one DM value, threads within a half warp (16 threads) will
access the input buffer in a quasi-fully coalesced manner. This also
applies for storing the result in the output buffer since all threads
within a row will shift by the same amount, resulting in stores which
are performed in a coalesced manner as well.

Shared memory is also used to reduce global memory reads. Each output
value requires $N_c$ additions, and performing these additions in
global memory would reduce performance drastically. To counter this,
each thread is assigned a cell in shared memory where the additions
are performed. The final result is then copied to global memory.

\subsection{Subband de-dispersion}
\label{subbandGpu}

\begin{figure} \centering
  \includegraphics[width=80mm]{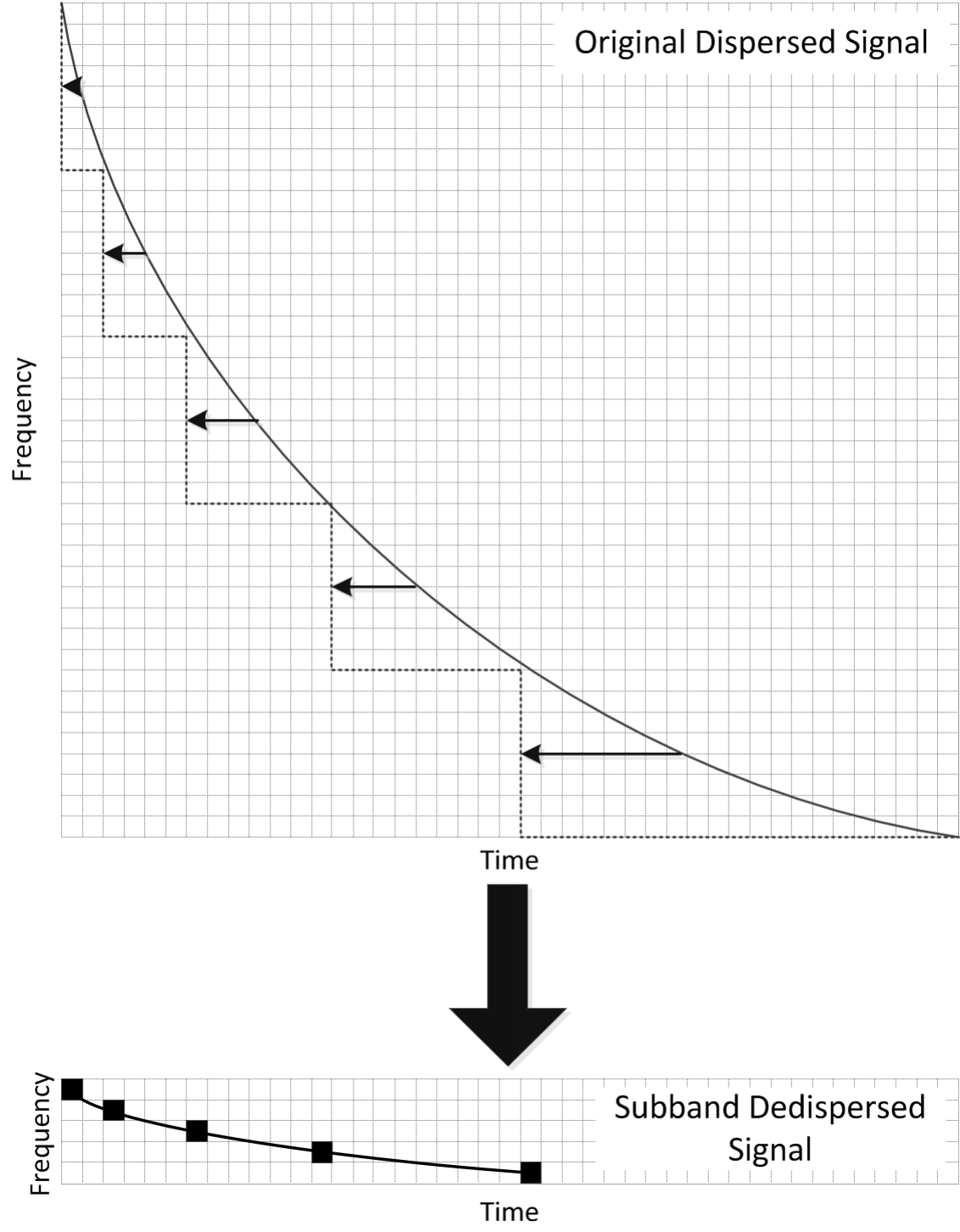} \caption[Subband
  de-dispersion illustration]
        {A simple illustration to visually depict how subband de-dispersion works. The channels are partitioned into a set
         of subbands where the delays corresponding to the nominal DM for every channel in the subband, minus the delay at the
	highest frequency in the subband, are subtracted from each channel. The subband de-dispersed signal is then
	further de-dispersed using normal brute-force de-dispersion for a range of DM values around the nominal DM value.}
\label{subbandDedispFigure}
\end{figure}

Subband de-dispersion uses aspects of brute-force de-dispersion,
however it is also influenced by the tree algorithm, which reuses sums
of groups of frequency channels for different DM values. It relies on
the fact that adjacent DM values (given an appropriate DM step) will
use overlapping samples during the summation, so it splits up the DM
range into several sub-ranges, each centred around a nominal DM value.
The bandwidth is also split into several subbands, resulting in a
partitioning of the set of channels. The delays corresponding to the
nominal DM for every channel in a subband, minus the delay at the
highest frequency in that subband, are subtracted from each subband
channel. This results in a partially de-dispersed set of subbands.
This scheme is depicted in figure \ref{subbandDedispFigure}. Normal
de-dispersion is then used to generate the de-dispersed time series
for the rest of the DM values within the same DM sub-range.

Depending on the number of subbands used, the size of the DM ranges,
as well as other factors, we can limit the error induced in the result
by these approximations. Further gains can be made by binning the
input samples when the dispersion is so high that a pulse with a
required width is smeared across multiple input samples. Binning
averages $N_b$ consecutive samples. To make our GPU code compatible
with the way PRESTO generates its survey parameters, binning has also
been implemented. There are, therefore, three main stages in subband
de-dispersion: \begin{enumerate}
 \item Perform data binning, if required
 \item Perform subband de-dispersion and generate the intermediary time series
 \item De-disperse the resultant time series to generate the final de-dispersed time series.
\end{enumerate}

\begin{figure}
\centering
\includegraphics[width=87mm]{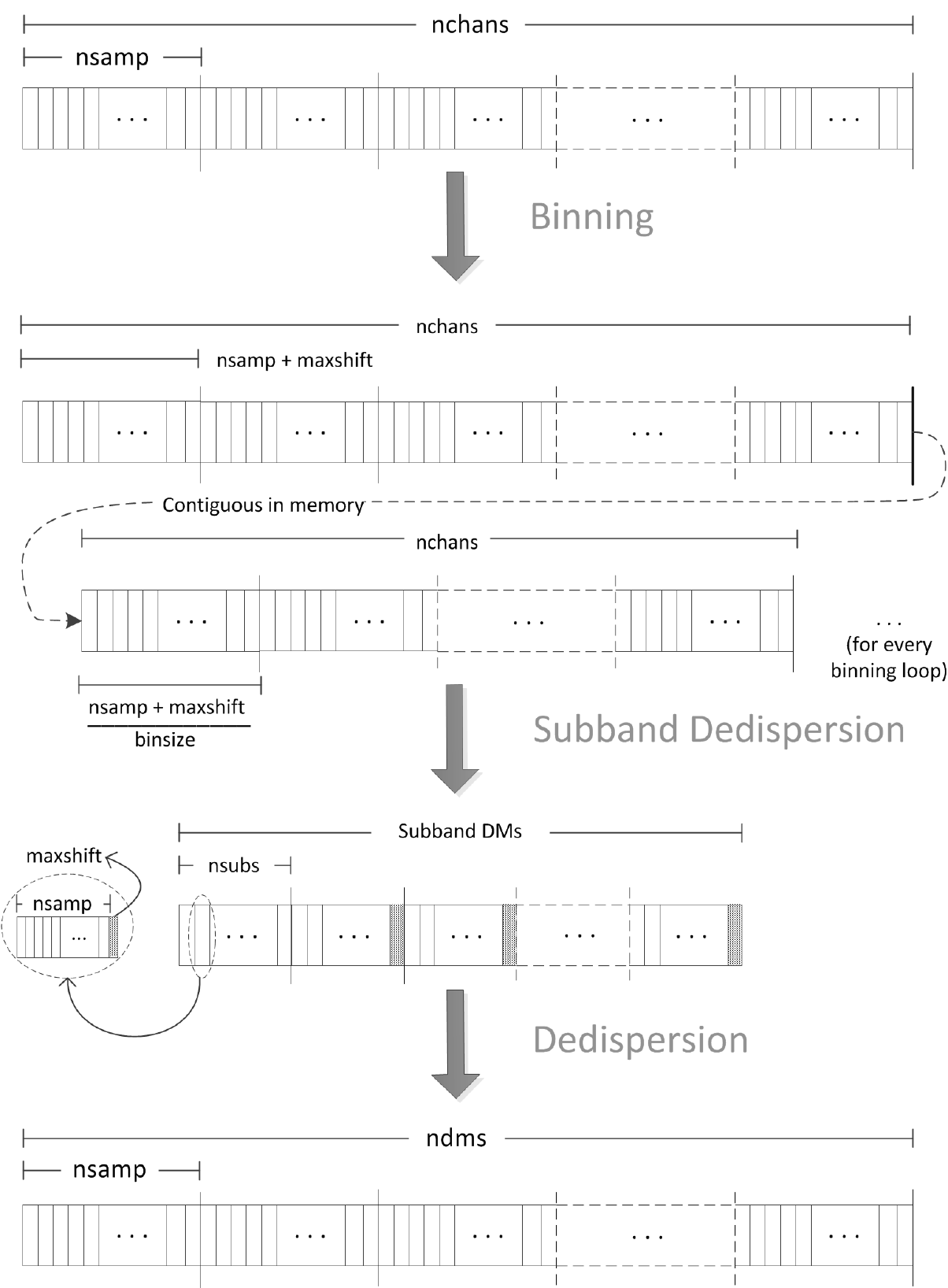}
\caption[Organisation in GPU memory during Subband De-dispersion]
        {Subband de-dispersion requires three passes of the data: binning, subband de-dispersion and brute force de-dispersion.
         During the entire process data is kept in GPU memory, and this illustration shows how this data is organized before
         and after each pass.}
\label{subbandMemoryFigure}
\end{figure}

Data transfers between the host and GPU are expensive, so the above
stages are performed in the GPU without any data going back to the
host, with appropriate data buffer re-organisation performed after
each kernel execution. The memory organisation after each stage is
depicted in figure \ref{subbandMemoryFigure}, which can be described
from top to bottom as follows:

\begin{enumerate}
\item The input buffer, containing $\left(n_{\rm samp} + m_{\rm
      shift}\right)\times n_{\rm chans}$ values
\item Binning averages $N_b$ adjacent samples, so the output of one
  binning loop will be
  $\left(n_{\rm samp} + m_{\rm shift}\right) \times n_{\rm chans}/N_b$, with each loop having a different value for $b$. The output
  of each binning loop is placed at the tail of the previous output,
  as shown in the diagram. For $l$ loops, the
  buffer will end up containing $l$ logical blocks, each with a different bin size $b_i$. The memory, in samples, required for this
  procedure is:
  \begin{equation}
    mem=\sum_{i=0}^l\left(\frac{\left(n_{\rm samp}+m_{\rm shift}\right) \times n_{\rm chans}}{b_i}\right)
  \end{equation}
\item Subband de-dispersion generates $N_{sub}$ intermediate
  time-series for each nominal DM, each consisting
  of $(n_{\rm samp} + m_{\rm shift})/b_i$ samples containing $n_{\rm subs}$ channels. Maxshift samples have to be preserved so that the 
  next stage can process all of $n_{\rm samp}$. The memory, in samples, required for this stage is :
  \begin{equation}
    mem=\sum_{i=0}^l\left(\frac{\left(n_{\rm samp}+m_{\rm shift}\right)\times n_{\rm subs} \times N_{sub}}{b_i}\right)
  \end{equation}
\item The final output consists of the de-dispersed time series for all the DM values, each containing $n_{\rm samp}/b_i$ 
  values. Thus the memory requirement for this stage, in samples, is:
  \begin{equation}
    mem = \sum_{i=0}^l\left(\frac{n_{\rm dms}\times n_{\rm samp}}{b_i}\right)
  \end{equation}
\end{enumerate}

Having defined the input and output memory requirements for all the
GPU stages, the number of samples which can be processed can then be
calculated. The size of the input and output buffers can be computed
by taking the size of the respective largest buffer from the
processing stages, which can then be used to compute the number of
samples which will fit in memory.

The subband de-dispersion kernel is very similar to the brute force
one, the only major change being that not all the channels are summed
up to generate the series, and more than one value is generated per
input sample. This makes the algorithm less compute intensive and more
memory limited (same number of input requests, more output requests).
However the number of nominal DM values is only a fraction of the
total number of DM values, and this greatly reduces the number of
calculations which need to be performed by the brute force algorithm.

\section{Results and comparisons} \label{resultsSection}

To test the code, a file containing a pulsed signal was generated
using the fake pulsar generator within SIGPROC\footnote{SIGPROC is 
a software package designed to standardize the analysis of various 
types of fast-sampled pulsar data} (Lorimer,
http://sigproc.sourceforge.net). The parameters which were used to
generate this fake file are listed in table \ref{surveyPlanTable}. The
fake filterbank data are generated as 1024 time-series, one for each
frequency channel. Each one is made up of a square pulse of height
$8\sqrt{1024} = 0.25$ and Gaussian noise with mean 0 and standard
deviation 1. The S/N of the average simulated pulse, integrated over
frequency has a mean value of 8. 

\begin{figure} \centering
  \includegraphics[width=80mm]{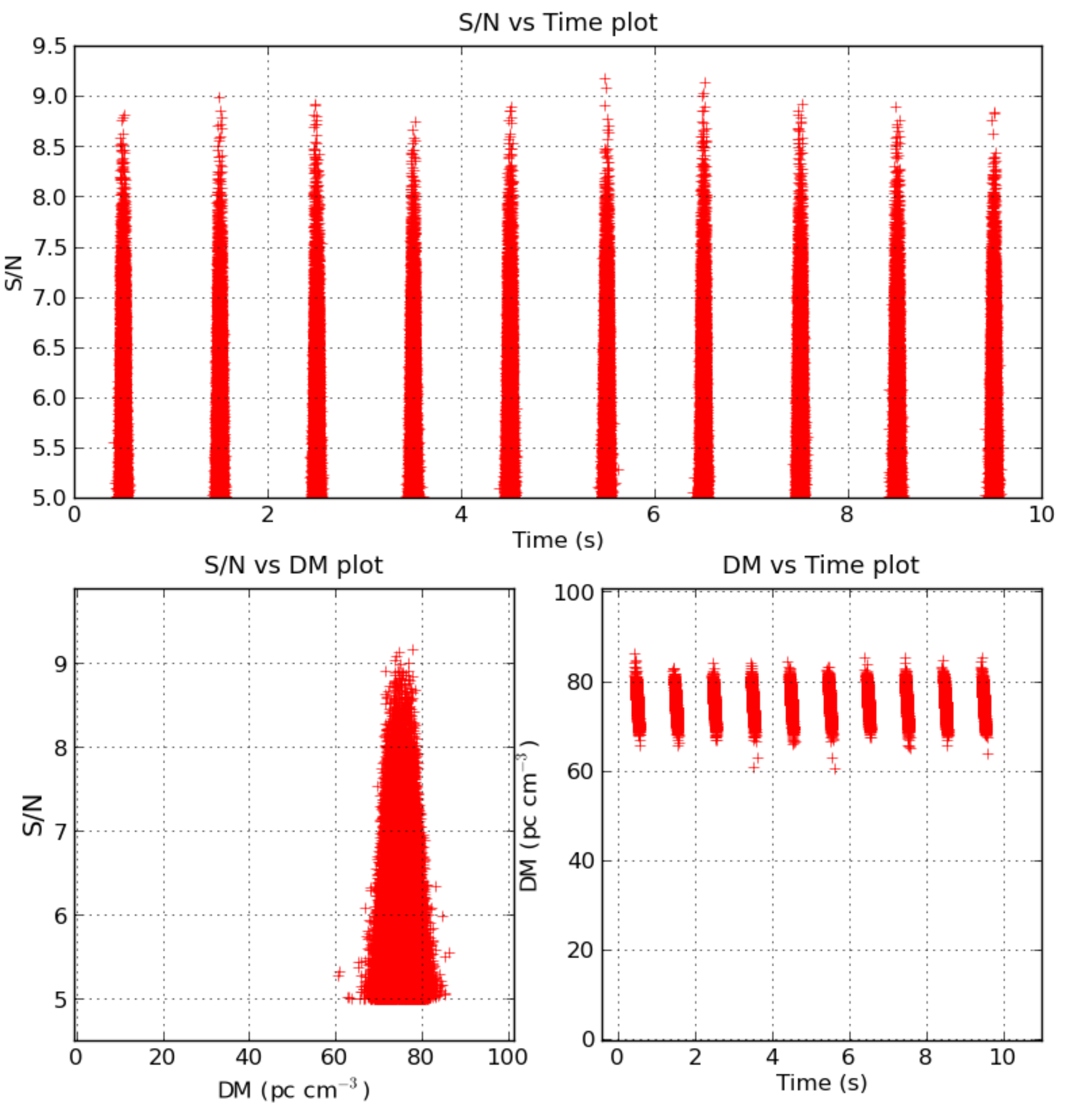} 
  \caption[Brute-force evaluation for a pulsar with an S/N of
                    5] {Brute-force de-dispersion output for an input
                    file containing a simulated pulsar (see text for
                    details). The plots are S/N versus time (top), S/N
                    versus DM (bottom left) and DM versus time (bottom
                    right).  A threshold of 5$\sigma$ is applied to
                    the output. The data points shown are at the DM of
                    the simulated pulsar, 75 $\text{pc cm}^{-3}$.}
  \label{brute2Figure}
\end{figure}

Brute-force de-dispersion using 1000 DM values with a DM step
of 0.1 $\text{pc cm}^{-3}$ was performed. Figure \ref{brute2Figure}
shows the output of the de-dispersion code, which captures all pulses
with S/N greater than 5.

\begin{table}
  \centering
  \begin{tabular}{l | c}
  Parameter          & Value \\
  \hline
  Pulsar Period      & 1000 ms \\
  Duty Cycle         & 1\%    \\
  Pulsar DM          & 75.00 $\text{pc cm}^{-3}$  \\
  Top Frequency      & 156.0 MHz  \\
  Channel Bandwidth  & 5.941 kHz  \\
  Number of Channels & 1024  \\
  Sampling Time      & 165 $\mu s$  \\
  \end{tabular}
  \caption{The parameters used to generate the fake file for
    evaluation. A pulsar with a period of 1s and 1\% duty cycle was
    created at a center frequency of 153 MHz with 6 MHz bandwidth. The
    bandwidth is divided into 1024 channels and a sampling time of 165
    $\mu$s was used. The DM is 75 $\text{pc cm}^{-3}$.}
  \label{surveyPlanTable}
\end{table}

\begin{figure}
  \centering \subfloat[Maximum number of
    samples]{\includegraphics[width=85mm]{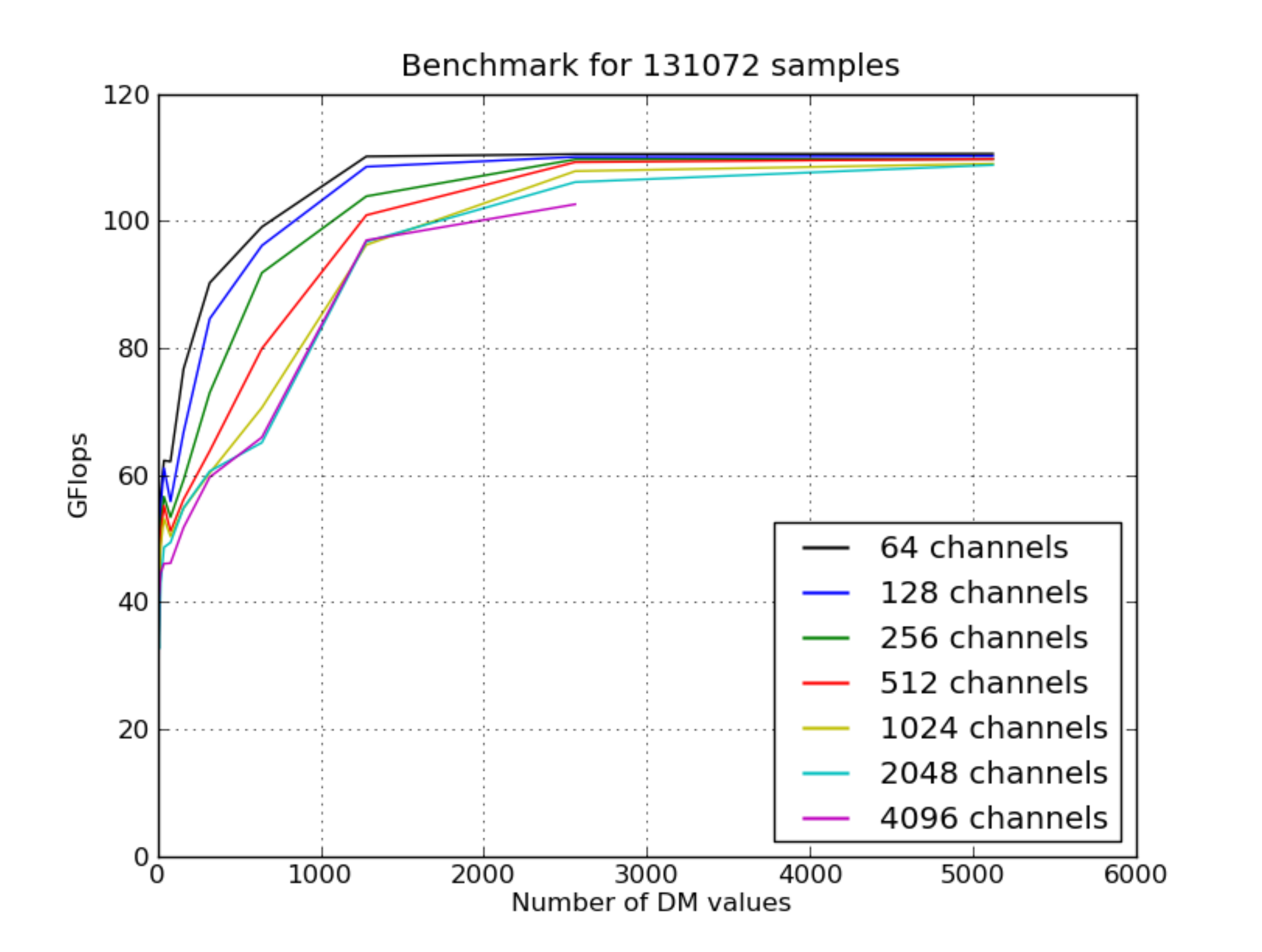}}
  \hspace{8mm}
  \subfloat[Maximum number DM values]{\includegraphics[width=85mm]{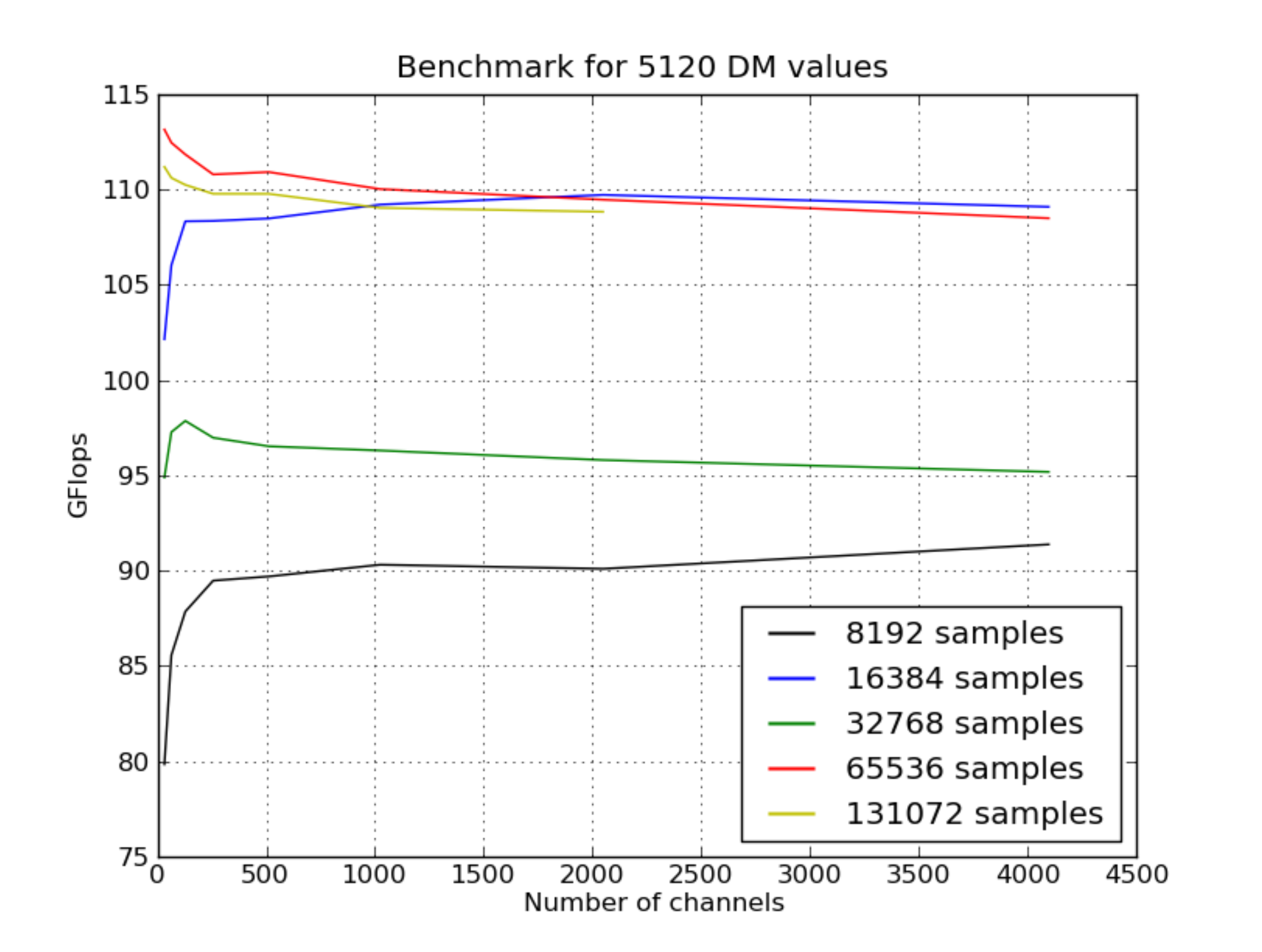}}
  \hspace{8mm}
  \subfloat[Maximum number of channels]{\includegraphics[width=85mm]{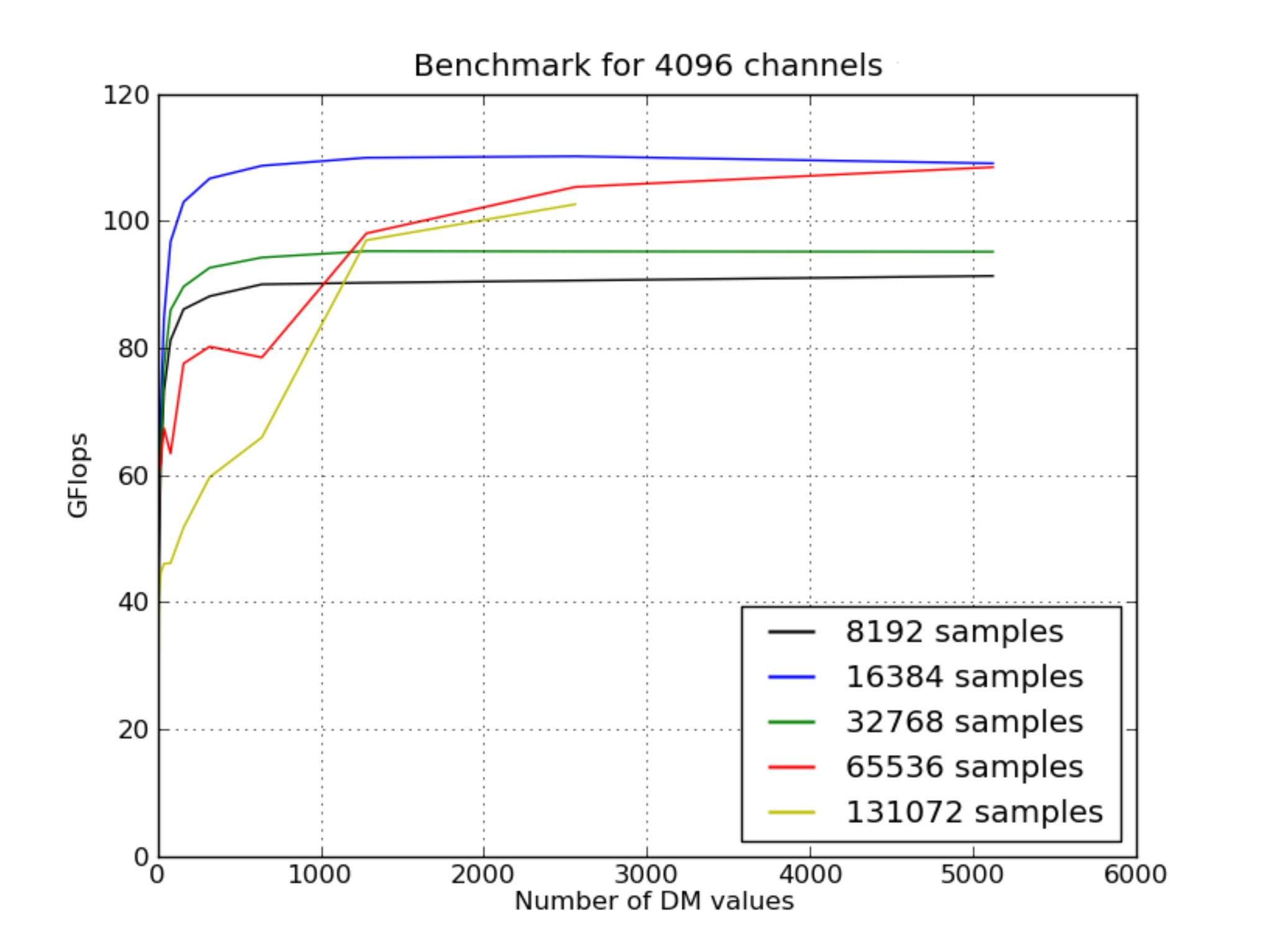}}
  \caption[CPU Performance plots]
  {Brute-force de-dispersion performance plots for a CUDA brute-force implementation. The general trend is for performance
    to increase linearly with increasing number of channels, samples and DM values, with different configurations reaching
    asymptotic behaviour at different peak performances. (Incomplete lines show cases where there was not enough memory 
    on the GPU to store the data required to perform the test.)}
  \label{timingGpuFigure}
\end{figure}

\begin{figure}
  \centering
  \subfloat[Maximum number of samples]{\includegraphics[width=85mm]{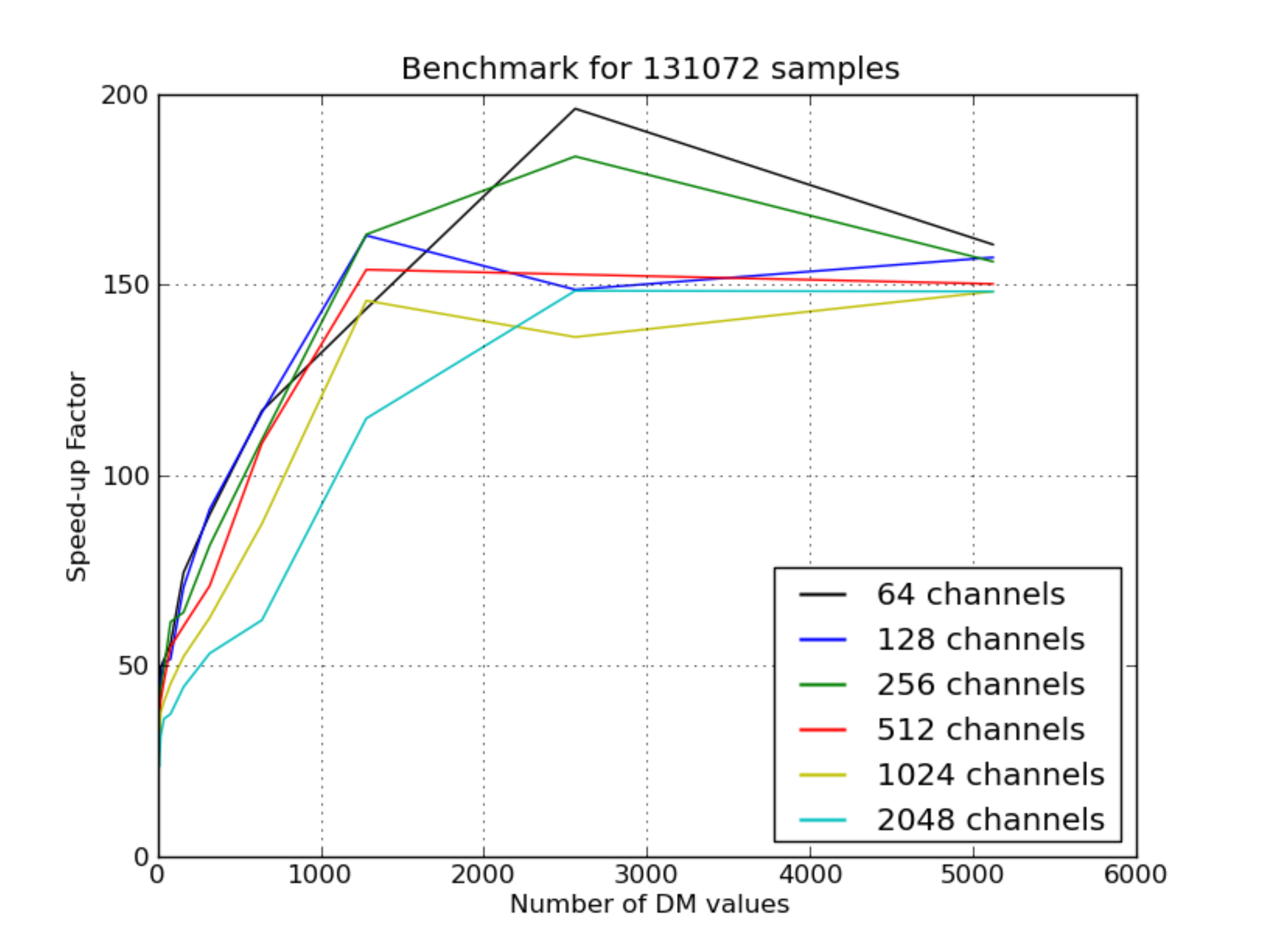}}            
  \hspace{8mm}
  \subfloat[Maximum number DM values]{\includegraphics[width=85mm]{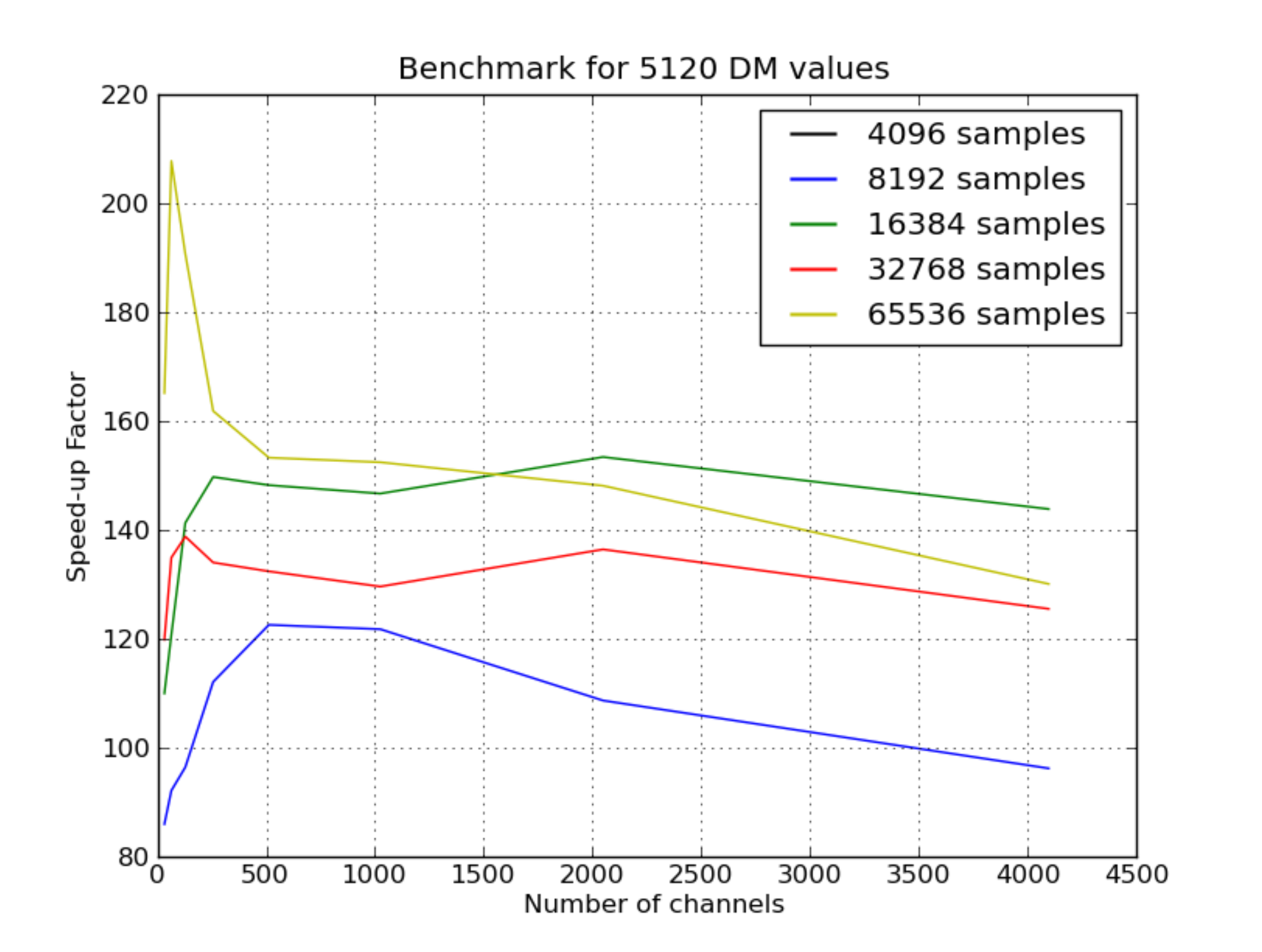}}
  \hspace{8mm}
  \subfloat[Maximum number of channels]{\includegraphics[width=85mm]{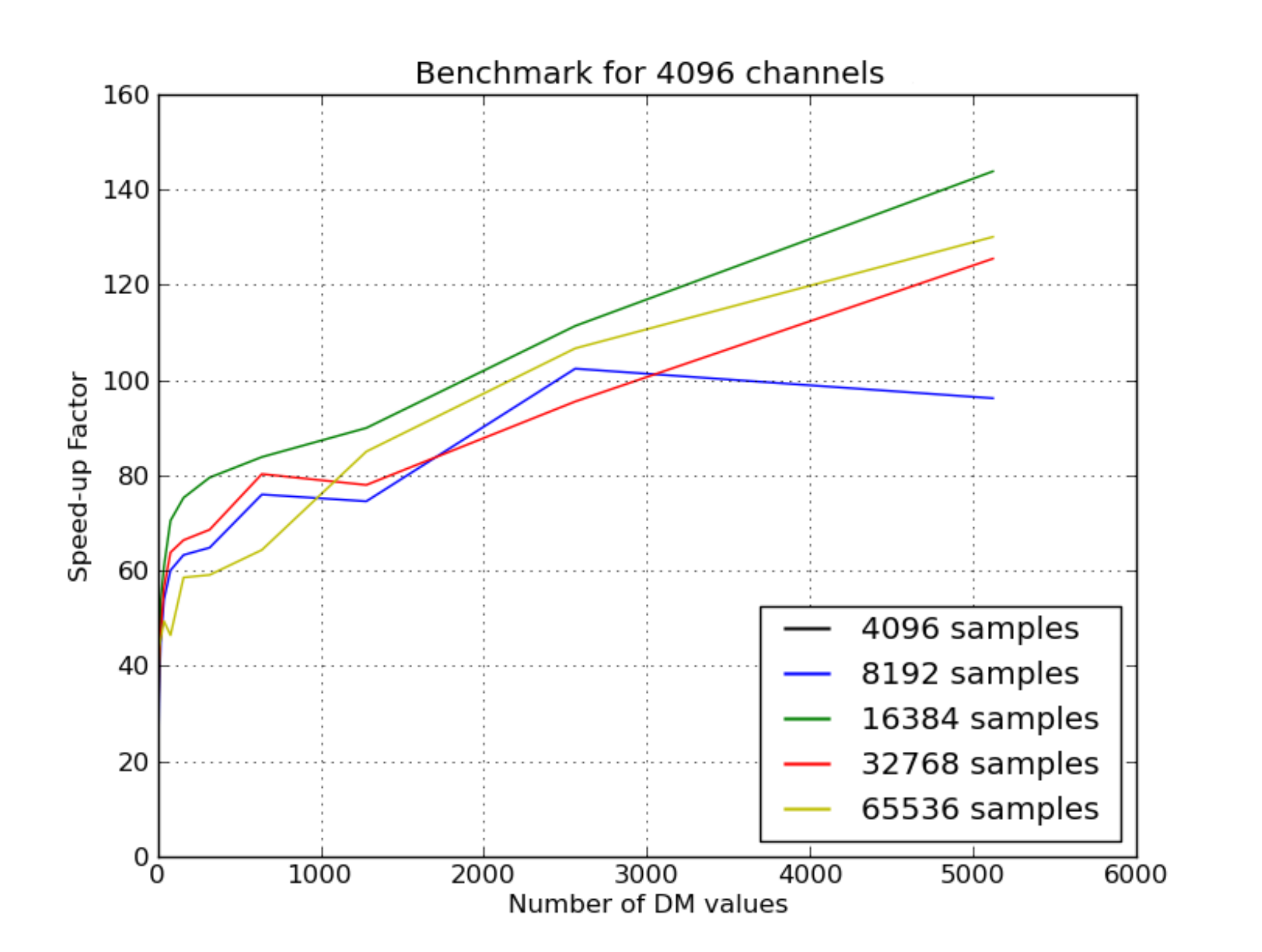}}
  \caption[GPU performance Speedup]
  {Brute-force de-dispersion speedup plots. For the maximum number of samples used in the tests (a), performance
    speedup converges to about 150$\times$, with peaks at different number of channels for different number of DM values. For the
    maximum number of DM values (b), the speedup decreases quasi-linearly with increasing number of channels due to maxshift
    offset. In the maximum number of channels case (c), performance increases quasi-linearly.}
  \label{bruteSpeedupFigure}
\end{figure}

\begin{figure}
  \centering
  \subfloat[The speed up of subband versus brute force
  de-dispersion]{\includegraphics[width=85mm]{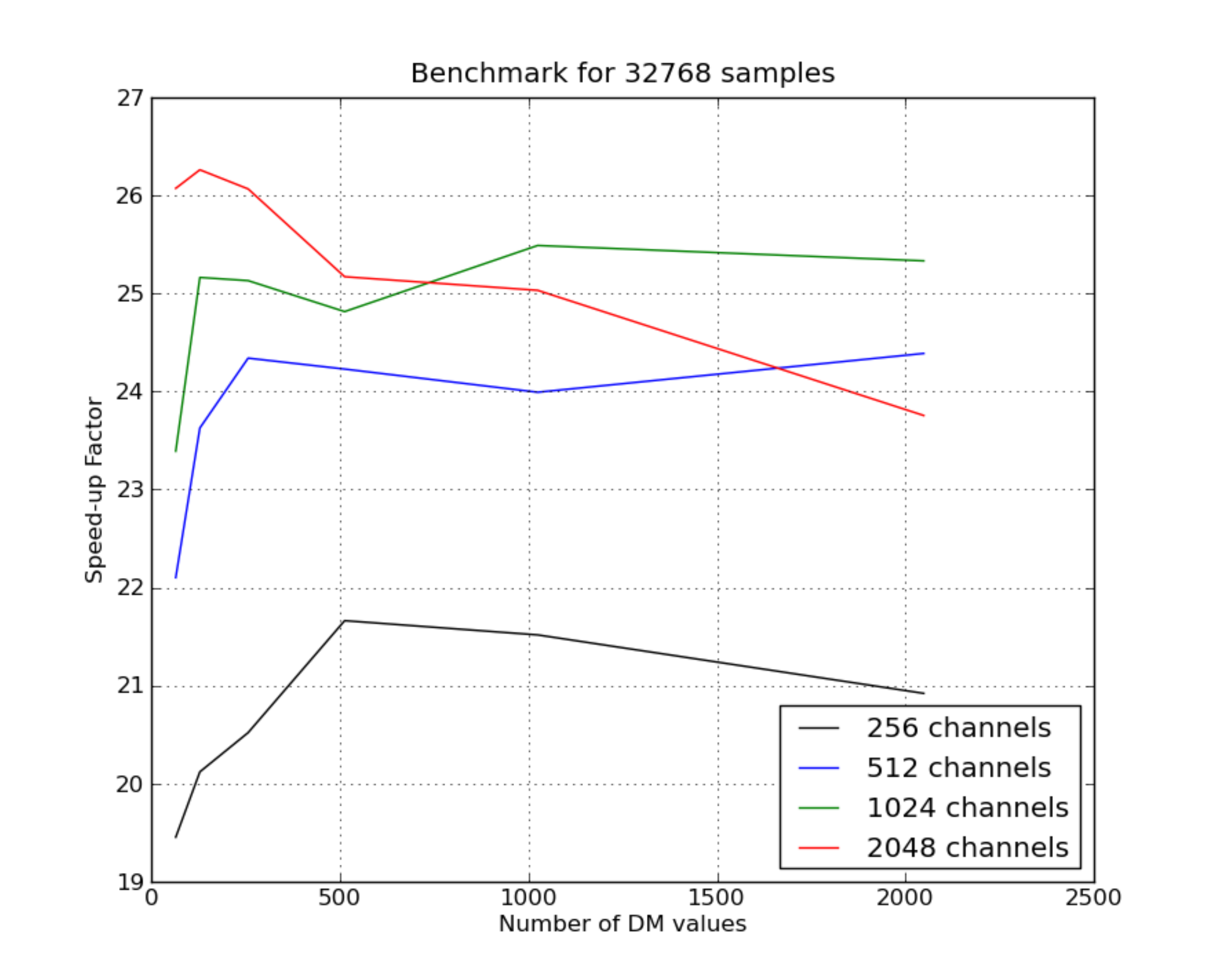}}
  \hspace{8mm}
  \subfloat[Different DM values per nominal DM in subband de-dispersion]{\includegraphics[width=85mm]{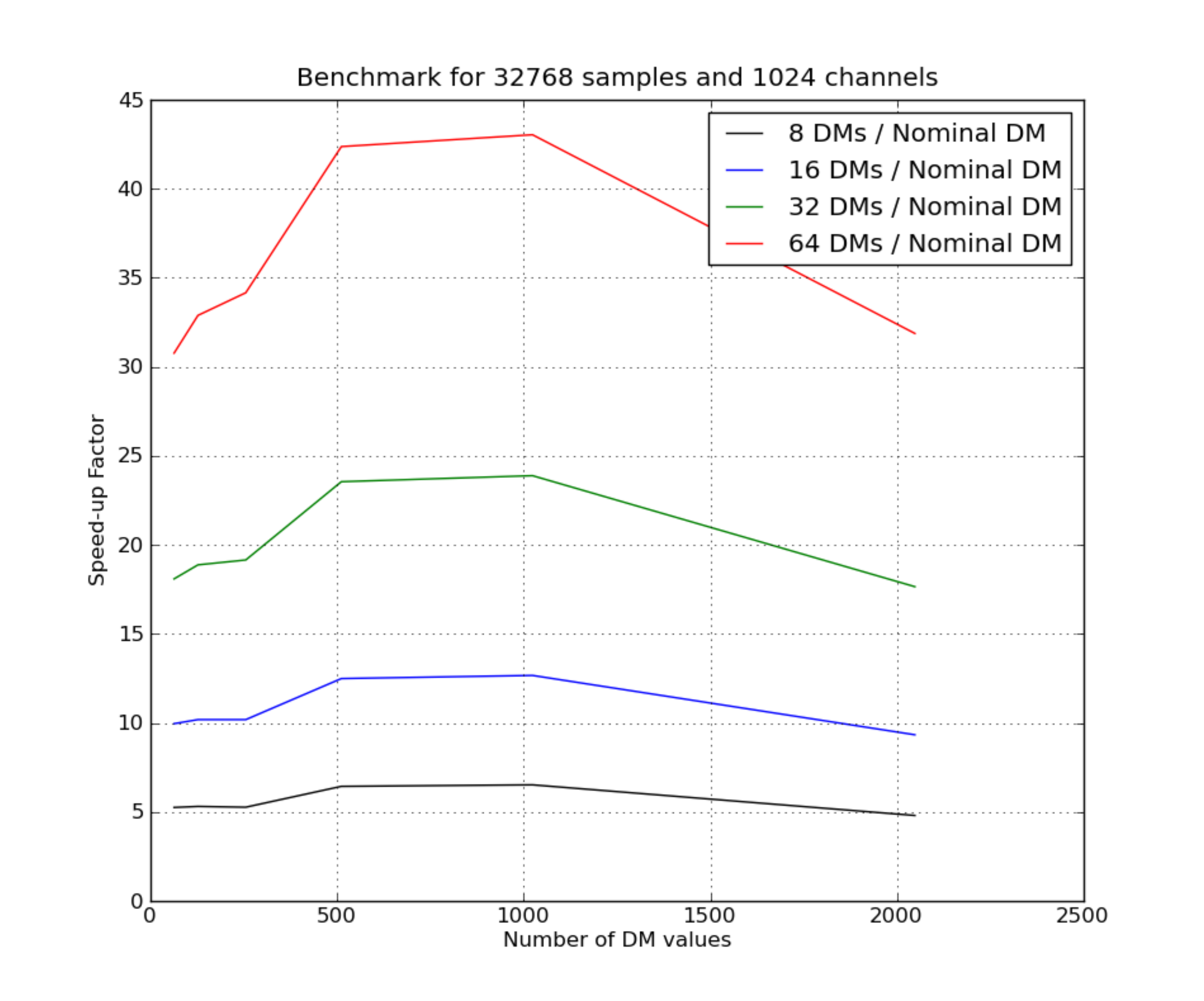}}
  \caption[De-Dispersion algorithms comparison plots]
  {The relative speed up of subband de-dispersion compared to brute
    force de-dispersion on the GPU. Plot (a) compares a series of
    de-dispersion runs with varying parameters using the two
    algorithms. The speed-up factor depends on optimal parameter
    combinations, the number of subbands, and the number of DM values
    per nominal DM used to split the DM-range, as shown in plot (b).
    See main text for further details.}
  \label{subbandComparisonFigure}
\end{figure}

The performance of the CUDA implementations has been measured. Fake
data is generated in the testing runs themselves, with all the
elements initialized to the same value. The time taken to generate and copy the data
to and from GPU memory is not included in the timings.

Figure \ref{timingGpuFigure} shows the performance achieved when
de-dispersing with different number of channels, samples and DM
values. Different parameter configurations will result in some
different optimal combinations, for example, in cases where the number
of data partitions to process is exactly divisible by the number of
processors and thread blocks being used. The general tendency is for
performance to increase linearly as the number of channels, samples
and DM values increases until the maximum GPU occupancy level is
reached, after which the behaviour becomes asymptotic. The optimal
block size is 128, since fewer threads will result in less latency
hiding and more threads will increase scheduling latency without
performance benefits. The grid size does not affect performance too
much, except for the case where there are too many threads in each
block.

As was already stated, the de-dispersion algorithm is memory-bound,
and both the GPU and CPU will spend most of their time waiting for
data. For this reason, the flop rate achieved on the GPU is a small
percentage of the theoretical peak for the C-1060, between 80 and 120
Gflops, which comes to about 15-20\%. The memory bandwidth achieved
within the GPU is about 55 GB/s, which is about 50\% of the
theoretical peak.

The same tests were performed on a CPU, specifically on one core of a 
QuadCore 
Intel Xeon 2.7 GHz. CPU-performance decreases quasi-linearly as the 
number of samples or channels increases due to cache misses. This 
performance is then compared with the appropriate GPU performance to 
produce the comparison plots in figure \ref{bruteSpeedupFigure}. This 
shows the speedup gained in performing brute force de-dispersion when 
using GPUs, for different parameter values. From these plots it follows 
that on average we get a speed of about \textit{two orders of magnitude},
between 50$\times$ - 200$\times$ depending on the parameters used, with the speedup
increasing as the number of input samples/channels increases.

The CUDA implementation was then compared to the two most commonly
used de-dispersion scripts, the one in PRESTO and the one in SIGPROC.
A fake file was generated containing a 600-second observation centered
at 153 MHz with a bandwidth of 6.24 MHz split into 1024 channels and
having a sampling rate of 165 ms (containing a total of about
$3.6\times10^6$ samples). This was run through the three software
suites for a \textit{single DM value}. For the GPU code and PRESTO only the
actual de-dispersion part was timed, whilst SIGPROC also loads from
file in the innermost loop so the timing contains some file I/O time
as well. The timings are listed in table \ref{bruteComparisonTable}.

\begin{table}
   \centering  
  \begin{tabular}{l | c}
  Suite     & Timing   \\
  \hline
  GPU Code      & 0.257s   \\
  PRESTO    & 28.321s  \\
  SIGPROC   & 58.099s  \\
  \end{tabular}
   \caption{Timing comparison between the GPU Code, PRESTO and SIGPROC for one DM. SIGPROC loads data in its innermost
           loop so some of the time listed is actually spent reading from file. The timing discrepancy between
           the GPU Code and PRESTO is consistent with the speedup plots.}
  \label{bruteComparisonTable}
\end{table}

The algorithm used to perform subband de-dispersion (both steps) is
almost identical to the one used in brute-force de-dispersion, so the
scaling tests were not repeated. Figure \ref{subbandComparisonFigure}
depicts a comparison plot between the two algorithms for various
de-dispersion parameters. The speed-up factor depends on optimal
parameter combinations as well as the number of subbands and number of
nominal DM values in the DM range employed for subband de-dispersion.
The speed-up factor decreases linearly with increasing number of
nominal DM values for a particular range, since more work needs to be
done in the first algorithm step (although the same number of DM
values are processed, the first step will generally be more intensive
since the data has not been reduced yet). The number of nominal DM
values and the number of subbands depend on the amount of dispersion
smearing permissible, where a large subband-DM step and few subbands
result in a higher amount of smearing. These values should be
fine-tuned to acquire the best balance between S/N and processing
speed.

The GPU subband de-dispersion implementation was also compared with
PRESTO's prepsubband script, on which the algorithm is based. A fake
file for a single-beam 60-second observation at 300 MHz with a
bandwidth of 16 MHz and 1024 channels was created. The plan used for
the test is listed in table \ref{subbandPlanTestTable}. The time taken
for the GPU code and PRESTO to process the entire file is \textit{90s and
  7540s respectively}. Again, this indicates that the GPU code is about two
orders of magnitude faster than the CPU implementation, in this case
84$\times$ faster. For this test, PRESTO was run in single-thread mode.

\begin{table}
  \centering
  \begin{tabular}{c | c*{6}{c}}
  Pass     & Low DM & High DM & $\Delta DM $ & Bin & $\Delta Sub_{DM}$ \\
           & ($\text{pc cm}^{-3}$) & ($\text{pc cm}^{-3}$) & ($\text{pc cm}^{-3}$) & & ($\text{pc cm}^{-3}$) \\
  \hline
  1        & 0.00   & 53.46   & 0.03         & 1   & 0.66              \\
  2        & 53.46  & 88.26   & 0.05         & 2   & 1.2               \\
  3        & 88.26  & 150.66  & 0.10         & 4   & 2.4               \\
  \end{tabular}
  \caption{Subband de-dispersion survey plan used to compare PRESTO and the GPU Code, for a 60-second observation at 300 MHz with a bandwidth of 16 MHz split across 1024 channels.}
  \label{subbandPlanTestTable}
\end{table}

\section{Real-time de-dispersion}
\label{realTimeSection}

\begin{figure}
\centering
\includegraphics[width=76mm]{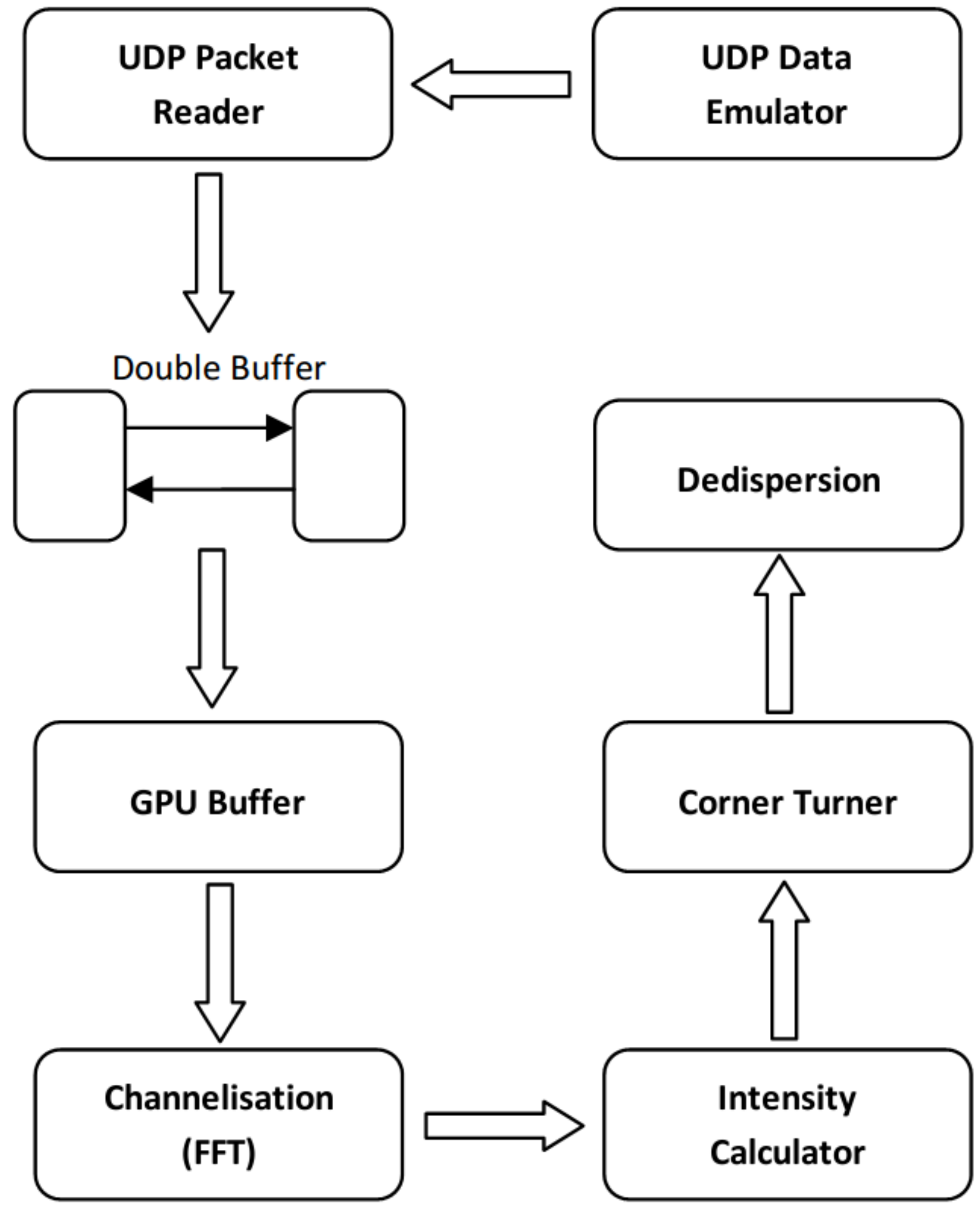}
\caption[Real-Time proof of concept] 
        {A setup for real-time de-dispersion. The UDP Data Emulator packetizes SIGPROC files and sends the data through UDP to
         the processing pipeline, where the packets are read, interpreted and stored in a double buffer. Once a buffer is full,
         it is forwarded to the GPU which performs channelisation, power calculation, a corner turn and de-dispersion.}
\label{realTimeFigure}
\end{figure}

The performance boost obtained from GPUs make them an ideal candidate
for use in real-time systems. An off-the-shelf server with a high-end
CUDA-enabled graphics card has enough power to de-disperse thousands
of DM values in real-time (depending on telescope parameters).
Additional features are required for such a system, such as a way to
read and interpret incoming telescope data, further channelisation and
buffering between the input stream and de-dispersion buffers.

As a proof of concept, the GPU code was extended to include a
channeliser (a simple FFT using the NVIDIA CUFFT library) and a kernel
to calculate the power from incoming complex voltages, to simulate
the situation of attaching such a machine to a baseband recorder. This
was used within a broader application which
\begin{inparaenum}
\item reads in UDP packets, filling up buffers within a double-buffer framework
\item forwards filled buffers to the GPU code
\item channelizes and calculates total power
\item transposes data so that it is in channel order
\item performs de-dispersion.
\end{inparaenum}
A UDP data emulator was used to create a simulated voltage stream from
SIGPROC fake data files and send them to the processing pipeline. This
setup is shown in figure \ref{realTimeFigure}.

A toy observation file was generated, whose parameters are defined in
table \ref{realTimeTable}, large enough so that multiple iterations of
the pipeline would be required, together with the processing
parameters. The brute-force de-dispersion algorithm was used for the
test. Note that the data emulator's output speed will not match the
simulated telescope's output data rate, so the way to determine
whether the pipeline is processing in real time is to time how long
the GPU takes to process one entire buffer, and then compare that with
the number of samples originally buffered.

\begin{table}
  \centering  
  \begin{tabular}{l | c}
  Parameter & Value \\
  \hline
  Center Frequency      & 610 MHz   \\
  Bandwidth             & 20 MHz  \\
  No. of Subbands       & 256  \\
  Sampling Time         & $12.8 \mu s$ \\
		        & \\
  Channels per Subband   &  8    \\
  Number of DM values    &  500   \\
  Maximum DM value      &  60 $\text{pc cm}^{-3}$  \\     
  \end{tabular}
   \caption{A toy observation for testing the real-time pipeline. A fake file was generated with an observation at a center
            frequency of 610 MHz and 20 Mhz bandwidth with 256 channels, producing 78125 spectra per second. The channelizer
            produces 8 channels per subband, and 500 DM values are used for the dispersion search, with a maximum DM of 
	    60 $\text{pc cm}^{-3}$.}
  \label{realTimeTable}
\end{table}

The GPU buffer sizes were set to $2^{19}$ spectra, equivalent to about
6.7s of telescope data, meaning that all the GPU processing for each
buffer must complete within this timeframe. The average timings for
each stage of the pipeline are listed in table \ref{timingsTable}. The
total processing time on an NVIDIA C1060 card is about 5.8s. This
leaves enough extra time for CPU-GPU synchronization and additional
memory operations, also providing enough leeway for the occasional CPU
processing burst due to other running processes or the OS itself. The
test was run on server with 2 QuadCore Intel Xeon 2.7 GHz and 12 GB
DDR3 RAM, which is a modest system for online processing.

\begin{table}
  \centering  
  \begin{tabular}{l | c}
  Stage & Time \\
  \hline
  CPU to GPU copy         & 475ms   \\
  Channelisation          & 458 ms  \\
  Intensity Calculation   & 20 ms  \\
  Corner Turn             & 112 ms  \\
  De-dispersion            & 4500 ms \\
  GPU to CPU copy         & 220 ms  \\
  \hline 
  Total                   & 5785 ms 
  \end{tabular}
   \caption{Timing for the several stages in the processing pipeline.}
  \label{timingsTable}
\end{table}

\section{Conclusions}
\label{conclusionSection}

We have implemented two de-dispersion algorithms, brute-force and
subband de-dispersion, using CUDA, which enables data-parallel
processing to be offloaded onto any number of connected CUDA-enabled
GPUs. This has led to a performance speedup of about two orders of
magnitude, between 50 and 200 for certain parameter configurations,
when compared to a single-threaded CPU implementation. Detailed
comparison with two traditional pulsar processing suites, PRESTO and
SIGPROC, confirm our results. Finally, a prototype for a real-time
dispersion search pipeline was designed, which reads in a UDP stream
of telescope data and performs FFT channelisation and de-dispersion.

Work is ongoing in this project, with plans to add several additional
processing modules in the pipeline. Coherent de-dispersion is useful
for studying pulsars whose DM value is already known. Other schemes
for de-dispersion are also being considered, such as performing chirp
analysis in the frequency domain to detect chirps representing
dispersed signals. GPUs provide us with enough processing power (per
unit cost) to be able to apply processing-intensive algorithms which
would otherwise be unfeasible on a conventional CPU system. As a
result, we are in a position to carry out real-time searches for
dispersed fast-transients with appropriate telescopes at low cost.

\section*{Acknowledgements}

The research work disclosed in this publication is partially funded by the 
Strategic Educational Pathways Scholarship (Malta). Aris Karastergiou is 
grateful to the Leverhulme Trust for financial support.

We thank the referee, Scott Ransom, for a constructive review of the paper.

\label{ack}

\label{lastpage}
\bibliographystyle{mn2e}
\bibliography{RealTimeDedispersion}

\begin{thebibliography}{13}
\expandafter\ifx\csname natexlab\endcsname\relax\def\natexlab#1{#1}\fi

\bibitem[{Ait~Allal {et~al}\mbox{.}(2009)Ait~Allal, Weber, Cognard, Desvignes,
  \& G.}]{Allal2009}
Ait~Allal D., Weber R., Cognard I., Desvignes G., G. T., 2009, in {Proceedings
  of 17th European Signal Processing Conference (EUSIPCO)}, pp. 2052--2056

\bibitem[{Barsdell, Barnes \& Fluke(2010)Barsdell, Barnes, \&
  Fluke}]{Barsdell2010}
Barsdell B., Barnes D., Fluke C., 2010, {Monthly Notices of the Royal
  Astronomical Society}, 408, 1936

\bibitem[{Bhat {et~al}\mbox{.}(2004)Bhat, Cordes, Camilo, Nice, \&
  Lorimer}]{Bhat2004}
Bhat N. D.~R., Cordes J.~M., Camilo F., Nice D.~J., Lorimer D.~R., 2004, ApJ,
  605, 759

\bibitem[{{Cordes}(2008)}]{cor08}
{Cordes} J.~M., 2008, in Astronomical Society of the Pacific Conference Series,
  Vol. 395, Frontiers of Astrophysics: A Celebration of NRAO's 50th
  Anniversary, {A.~H.~Bridle, J.~J.~Condon, \& G.~C.~Hunt}, ed., pp. 225--+

\bibitem[{{Cordes} \& {McLaughlin}(2003)}]{cm04}
{Cordes} J.~M., {McLaughlin} M.~A., 2003, ApJ, 596, 1142

\bibitem[{Dodson {et~al}\mbox{.}(2010)Dodson, Harris, Pal, \&
  Wayth}]{Dodson2010}
Dodson R., Harris C., Pal S., Wayth R., 2010, in {Proceedings of Science,
  ISKAF2010 Science Meeting}

\bibitem[{Lorimer {et~al}\mbox{.}(2007)Lorimer, Bailes, McLaughlin, Narkevic,
  \& F.}]{lorimer2007}
Lorimer D.~R., Bailes M., McLaughlin M.~A., Narkevic D.~J., F. C., 2007, Sci,
  318, 777

\bibitem[{Lorimer \& Kramer(2005)}]{LorimerKramer2005}
Lorimer D.~R., Kramer M., 2005, {Handbook of Pulsar Astronomy}. {Cambridge
  University Press}

\bibitem[{{Macquart, J. P. and Bailes M., et al}(2010)}]{Macquart2010}
{Macquart, J. P. and Bailes M., et al}, 2010, {Publications of the Astronomical
  Society of Australia}, 27, 272

\bibitem[{{McLaughlin} {et~al}\mbox{.}(2006){McLaughlin}, {Lyne}, {Lorimer},
  {Kramer}, {Faulkner}, {Manchester}, {Cordes}, {Camilo}, {Possenti}, {Stairs},
  {Hobbs}, {D'Amico}, {Burgay}, \& {O'Brien}}]{mll+06}
{McLaughlin} M.~A. {et~al.}, 2006, Nature, 439, 817

\bibitem[{{NVIDIA Corporation}(2010)}]{cudaOnline}
{NVIDIA Corporation}, 2010, {CUDA Zone}.
  \url{http://www.nvidia.com/object/cuda_home_new.html}

\bibitem[{{Ransom S.}(2001)}]{Ransom2001}
{Ransom S.}, 2001, PhD thesis, Harvard Univserity, Cambridge, Massachusetts

\bibitem[{van Straten \& Bailes(2011)}]{VanStraten2010}
van Straten W., Bailes M., 2011, Publications of the Astronomical Society of
  Australia, 28, 1

\end{thebibliography}

\end{document}